\documentclass[showpacs,superbib,10pt,preprint,final,twocolumn,aps]{revtex4}%
\usepackage{amsfonts}
\usepackage{amsmath}
\usepackage{amssymb}
\usepackage{graphicx}%
\begin{document}
\input epsf.sty \flushbottom

\title{Nonlinear current-induced forces in Si atomic wires}
\author{Zhongqin Yang}
\affiliation{Department of Physics, Virginia Polytechnic Institute and State University,
Blacksburg, Virginia 24061}
\author{Massimiliano Di Ventra\cite{MD}}
\affiliation{Department of Physics, Virginia Polytechnic Institute and State University,
Blacksburg, Virginia 24061}

\begin{abstract}
We report first-principles calculations of current-induced forces in Si atomic wires 
as a function of bias and wire length. We find that these forces are strongly 
nonlinear as a function of bias due to the competition between 
the force originating from the scattering states and the force due to bound states. 
We also find that the average force in the wire is larger the shorter the wire, suggesting 
that atomic wires are more difficult to break under current flow with increasing length. The last finding is in agreement with recent experimental data. 

\end{abstract}

\maketitle

Current-induced atomic motion (electromigration) has always been of concern in 
microelectronics since it was found to be a major failor mechanism in aluminum 
conductors.~\cite{blech} Such an effect is 
more pronounced the smaller the material dimensions as it is the case, for instance, 
for metallization lines in conventional circuits.~\cite{hauder}  It does not come as 
a surprise then that this effect has recently attracted considerable interest in 
nanoscience. In particular, current-induced forces in atomic~\cite{lango,agr,ita,yas,tod1} and molecular 
wires~\cite{aya,md1} have been investigated. These systems show interesting 
physical properties which make them promising candidates for future electronic 
applications.~\cite{avir} It was found  
that current-induced forces in a nanoscale wire can distort its local atomic structure and affect its stability according to the external 
bias and the wire length.~\cite{agr,ita,yas,tod1,md1,miochem} Longer atomic wires, for instance, have been found to be more stable with respect to current-induced forces than shorter 
wires.~\cite{yan}

Over the years considerable theoretical and experimental  
work has been devoted to understand the microscopic origin of electromigration.~\cite{solid} 
It has become clear, for instance, that this phenomenon depends strongly on the microscopic details of the self-consistent electric field that is created upon scattering of the electrons across the region of interest. Despite the progress in the fundamental understanding of current-induced forces, however, 
it is still unclear what determines, e.g., the resistance of atomic and 
molecular wires to electromigration effects or what is the dependence of these forces 
on the external bias and wire length. 

In this letter we explore the above issues using first-principles approaches. 
In particular, we study current-induced forces as a function of bias in Si atomic wires of 
varying length (we consider wire lengths of up to four Si atoms). 
We find that current-induced forces are strongly 
nonlinear as a function of bias due to the competition between 
the force originating from the continuum of states and the force due to states in the 
discrete part of the spectrum. 
The force originating from the bare electrodes (we will call this force direct~\cite{prec}) varies  
almost linearly as a function of bias. 
We also find that the average force in the wire is larger the shorter the wire, suggesting 
that atomic wires of increasing length are more difficult to break under current flow. 
The last finding is in agreement with recent experimental data.~\cite{yan}  

A schematic of the system investigated is depicted in the inset of Fig.~\ref{fig1}. It 
consists of a Si atomic wire sandwiched between two gold surfaces that we model with ideal 
metals (jellium model).~\cite{langt,md2} The interior electron density of the 
electrodes is taken equal to the value for metallic gold ($r_s\approx 3$). 
The electric current is calculated using the method described in 
Refs.~\onlinecite{langt,md2}. 
The electron wave functions are computed by solving 
the Lippman-Schwinger 
equation iteratively to self-consistency in steady state. 
Exchange and correlation are included in the density-functional 
formalism within the local-density 
approximation.~\cite{langt,md2} The 
current is computed from the wave functions $|\psi\rangle$ of the electrode-molecule system. 
The force ${\bf F}$ acting on a given atom at position ${\bf R}$
due to the electron distribution as modified by the external bias is given by the 
Hellmann-Feynman-type of theorem developed in Ref.~\onlinecite{md1}:~\cite{prec1}
\begin{equation}
{\bf F}= -\sum_i \langle\psi_i\left|\frac{\partial H}{\partial
{\bf R}}\right|\psi_i\rangle -
\lim_{{\Delta}\rightarrow 0} \int_{\sigma}dE
\langle\psi_{\Delta}\left|\frac{\partial H}{\partial {\bf R}}\right|
\psi_{\Delta}\rangle.
\label{equation}
\end{equation}
The sum and integral in Eq. (\ref{equation}) include spin variables also. The first term 
on the RHS of Eq. (\ref{equation}) is the usual Hellmann-Feynman contribution to the force 
due to localized electronic states $|\psi_i\rangle$. The second term is the contribution to the force due to 
the continuum of states.~\cite{md1} It is calculated by constructing, for each 
energy in the continuum, square-integrable wavefunctions 
$|\psi_{\Delta}\rangle$ in an energy region $\Delta$ 
\begin{equation}
|\psi_{\Delta}\rangle={\cal A}\int_{\Delta}dE\psi,
\label{newpsi}
\end{equation}
where ${\cal A}$ is a normalization constant and the 
$\psi$'s are single-particle wavefunctions in the continuum, solutions of the 
Lippmann-Schwinger equation.~\cite{md1}
The continuum integration $\sigma$ covers the part of the spectrum occupied by the
electrons at a given bias. Finally, the total force on the atom includes a trivial ion-ion interaction.  

We start the calculations by first relaxing the atomic positions at zero bias. 
For all different wire lengths the relaxed Si-jellium surface bond length is about 2 a.u., 
and the relaxed Si-Si bond distance is about 4.2 a.u. In Fig.~\ref{fig1}(a),(b),(c) we plot the total force as a function of bias on each Si atom for wires composed of two, three and four Si atoms, respectively. The atomic positions have been fixed at the equilibrium position at zero bias, 
and the atoms are 
labeled with increasing number starting from the closer to the left electrode 
(see Fig.~\ref{fig1}). 
In Fig.~\ref{fig1}(d) we plot the average force (sum of the forces on each atom 
divided by the number of atoms in the wire) as a function of bias and the force 
for a wire composed of a single Si atom. Positive force pushes the atom to the right, i.e., 
opposite to electron flow (the left electrode is positively biased).   
It is clear from Fig.~\ref{fig1}(a),(b),(c) that the 
force on atoms is a nonlinear function of the bias (exept for the single-atom 
wire, see below and Ref.~\onlinecite{prec3}) while the average force 
Fig.~\ref{fig1}(d) saturates with 
the number of atoms in the wire. We first discuss the nonlinear behavior of the forces and 
then discuss the average force.

For a small external bias (0.01 $V$, i.e., linear-response regime), the
current-induced forces in the 3-Si wire satisfy the zero-force sum rule 
while for the 2-Si and 4-Si wire 
all atoms are pushed opposite to the current flow with almost equal force 
as it can be inferred from symmetry 
considerations~\cite{tod1}. On the other hand, strong 
nonlinearities in the current-induced forces appear at voltages above 0.1 $V$. 
For instance, in the 2-Si wire the 
Si atom closer to the left electrode (Si$_1$ in the inset of Fig.~\ref{fig1}) is pushed against 
the electron flow for biases less than 0.5 $V$ and moves along the 
electron flow for larger biases. On the other hand, the second Si atom (Si$_2$ in the 
inset of Fig.~\ref{fig1}) is pushed 
against the electron flow at all external voltages. 
Similar effects, involving different atoms in the wire, occur at even smaller 
voltages for the 4-Si wire. In the same vein, 
for the 3-Si wire the zero-force sum rule is not satisfied already at 0.1 $V$. It is 
interesting to note that for large biases the largest force occurs on the second Si atom 
from the left. The break-up of the wires is thus likely to nucleate from the bonds of this 
atom. In 
general, each atom experiences a force due to current flow that is nonlinear in the external 
bias. Such nonlinearities affect the atomic redistribution in the wire and eventually 
its resistance to current-induced rupture. 
In order to understand this nonlinear behavior, we study the different 
contributions to the forces as a function of bias: 
(a) the contribution from the electrodes without the atoms in between (direct 
force~\cite{solid}), 
(b) the contribution from the continuum part of the spectrum, and (c) the part of the force 
originating from the discrete spectrum. The force originating from the ion-ion interaction 
does not depend on the bias and we thus not discuss it here. For the sake of simplicity we 
discuss the 3-Si wire case. Similar considerations are valid for 
other wire lengths. We plot in Fig.~\ref{fig2} the three different 
contributions considered. As expected, the direct force is almost linear with the bias. Small deviations 
from linearity appear due to the small deviations from linear decay of the electrostatic 
potential close to the electrode surfaces.~\cite{langt,md2} Furthermore, due to this 
deviation, the force on the central atom in the wire is larger than the force on the two 
atoms close to the electrode surfaces at any bias (see Fig.~\ref{fig2}). The latter atoms also 
experience a force of similar magnitude with deviations occurring at large voltages. 
From Fig.~\ref{fig2} is also evident that for biases larger than 0.5 $V$ the major contribution to the force on each atom comes 
from the direct force. The nonlinearities in the total force then 
orginate from the competition between the force due the states in the 
continuum (scattering states) and the force due to the bound states (in this 
case, those states below the band bottom of the left electrode). The force on the central atom of the wire 
(Si$_2$) due to the scattering states pushes the atom to the left, i.e. along with the electron flow. This 
force is almost linear with the bias. The force on the same atom due to the discrete spectrum, on the other hand 
is almost zero even at very large voltages. This behavior can be understood by looking at the extra charge 
localized on the atom at any given bias. This quantity can be estimated by integrating the charge around a given atom 
in a box centered on that atom with one face parallel to the electrode surfaces and whose size in the perpendicular direction is equal to the bond distance between the atoms. The difference 
between this charge at zero bias and the corresponding charge at any given bias is plotted in Fig.~\ref{fig3} for the 
three different Si atoms. Apart from small fluctuations as a function of bias, the extra charge on the central 
atom is zero. The charge on this atom is thus practically constant at any bias. 
Since, in this system, the number of bound states does not change with bias, their 
contribution to the force is constant [see Fig.~\ref{fig2}(c)]. On the other hand, the force 
due to the continuum [second term on the RHS of Eq. (1)], increases linearly with bias due to the larger energy 
integration in Eq. (1) for larger voltages.~\cite{prec3} 

The continuum and discrete contribution to the force is nonlinear for the atoms closer to the electrodes (Si$_1$ and 
Si$_3$ of Fig.~\ref{fig2}). For the Si$_1$ atom, 
for instance, the continuum force is almost constant up to about 1 $V$, while the force from the discrete spectrum 
increases in magnitude with bias in the same voltage range (the sign of the force corresponds to the atom pushed along 
with the current flow). For biases above 1 $V$, the continuum force increases in magnitude and the force due 
to bound states is almost constant. An opposite trend is observed for the Si$_3$ atom (see Fig.~\ref{fig2}). 
The opposite trend can be explained again by looking at the extra charge on these atoms (Fig.~\ref{fig3}). 
This charge is of similar magnitude but of opposite 
sign indicating that there is a charge transfer from the right electrode to the left electrode when current flows with 
consequent creation of a local electric dipole.~\cite{prec2} 

While it is difficult to extract general trends in the overall resistance of the wires 
to electromigration by looking at the forces on each atom, the average force follows 
quite a simple trend as a function of bias and wire length. In Fig.~\ref{fig1}(d) we plot 
such a quantity for wires of two, three and four Si atoms. For comparison we also plot the 
force for a single Si atom. The average force is almost linear as a function of bias even 
for large biases (deviations occur above 2.5 $V$ for the 2-Si wire). Furthermore, the 
force is larger the smaller the wire length and, in particular, it is almost equal for the 
3- and 4-Si wires.~\cite{prec4} This trend indicates that short wires are easier to break than longer 
wires, and the average force reaches a ``bulk'' value at wire lengths of 
only three atoms, i.e., wire lenghts of less than 10\AA. Similar trends 
have been observed in experiments with Au atomic wires.~\cite{yan} 

We conclude this paper by studying the effects of current-induced atomic 
relaxations on the current-voltage (I-V) characteristics of atomic wires. It was found 
by Di Ventra {\em et al.}~\cite{md1,miochem} that current-induced atomic relaxations 
in molecular wires do not substantially affect the absolute value of the current for 
large voltages and current densities. We find that this trend is also valid in the 
present case of atomic wires, suggesting it is a general trend valid for both atomic and molecular wires when current flow is mainly coherent. 
This is illustrated in Fig.~\ref{fig4} where we plot the 
I-V characteristics for the 3-Si wire with and without current-induced atomic relaxations. 
In the inset of the same figure we show the relaxed atomic positions for 
each atom at selected biases. It is 
clear from Fig.~\ref{fig4} that small changes in the absolute value of the 
current when the atoms are relaxed are observed even at large biases. 
Similar results are valid for the other wires as well. 

In conclusion, we have reported first-principles calculations of current-induced forces 
in Si atomic wires. We find that these forces are generally strongly 
nonlinear as a function of bias. Since the direct force from the bare electrodes is almost 
linear as a function of bias, the nonlinearity originates from the competition between 
the scattering-state and discrete-spectrum force. 
We also find that the average force in the wire is larger the shorter the wire, suggesting 
that atomic wires are more difficult to break under current flow with increasing length. 
Finally, current-induced relaxations are found to change only slightly the absolute 
value of the current even at large voltages. A similar effect has been predicted for 
molecular wires, indicating it is a general trend for nanoscale wires 
when coherent scattering is the main transport mechanism. 

We thank Y.-C. Chen for useful discussions. This work is supported in part by the National Science Foundation 
Grants No.
DMR-01-02277 and DMR-01-33075, Carilion Biomedical Institute, and Oak Ridge
Associated Universities. Acknowledgement is also made to the Donors of The
Petroleum Research Fund, administered by the American Chemical Society, for
partial support of this research. The calculations reported in this paper were
performed on the beowulf cluster of the Laboratory for Advanced Scientific
Computing and Applications at Virginia Tech.

\newpage

\begin{figure}
%\hspace*{0cm} \epsfxsize10cm \epsfbox{fig1Siforce.ps}
\vspace*{0.5cm}
\caption{Total force as a function of bias in atomic wires containing (a) two, (b) three,
and (c) four Si atoms. The inset shows a schematic of one of the wires investigated. 
The atoms are labeled with increasing number from the left electrode (see inset). (d) The average force in the Si wires. 
The left electrode is positively biased.}
\label{fig1}
\end{figure}

\begin{figure}
%\hspace*{0cm} \epsfxsize10cm \epsfbox{fig2Siforce.ps}
\vspace*{0.5cm}
\caption{Different contributions to the total force for a 3-Si wire: (a) direct force (see text), (b) 
scattering-state contribution, and (c) contribution from the discrete spectrum.}
\label{fig2}
\end{figure}

\begin{figure}
%\hspace*{0cm} \epsfxsize10cm \epsfbox{fig3Siforce.ps}
\vspace*{0.5cm}
\caption{Extra charge (see text) on the three Si atoms in the 3-Si
wire as a function of bias.}
\label{fig3}
\end{figure}

\begin{figure}
%\hspace*{0cm} \epsfxsize10cm \epsfbox{fig4Siforce.ps}
\vspace*{0.5cm}
\caption{I-V curve of the 3-Si wire with and
without the effect of current-induced atomic relaxations. The inset shows 
the unrelaxed (open circles) and relaxed (full circles) atomic 
positions of the Si atoms between the two electrodes (vertical thin rectangles) 
at selected biases. The atoms move opposite to the electron flow.}
\label{fig4}
\end{figure}


\begin{thebibliography}{99}                                                                                              
\bibitem[*]{MD}E-mail address: diventra@vt.edu.

\bibitem {blech}See e.g., I. A. Blech and H. Sello, in \textit{Physics of Failure in
Electronics}, edited by T. S. Shilliday and J. Vaccaro (USAF, Rome Air
Development Center, 1967), Vol. \textbf{5}, p. 496.

\bibitem{hauder} M. Hauder, J. Gstottner, W. Hansch, and D. Schmitt-Landsiedel, 
Appl. Phys. Lett. {\bf 78}, 838 (2001). 

\bibitem{lango} N.D. Lang, Phys. Rev. B {\bf 45}, 13599 (1992); {\bf 49}, 2067 (1994). 

\bibitem {agr}N. Agra\"{\i}t, C. Untiedt, G. Rubio-Bollinger, and S. Vieira,
Phys. Rev. Lett, \textbf{88}, 216803 (2002).

\bibitem{ita} K. Itakura, K. Yuki, S. Kurokawa, H. Yasuda, and 
A. Sakai, Phys. Rev. B {\bf 60}, 11163 (1999). 

\bibitem {yas}H. Yasuda and A. Sakai, Phys. Rev. B \textbf{56}, 1069 (1997).

\bibitem {tod1}T. N. Todorov, J. Hoekstra, and A. P. Sutton, Philos. Mag. B
\textbf{80}, 421 (2000); T. N. Todorov, J. Hoekstra, and A. P. Sutton, Phys. Rev. Lett.
\textbf{86}, 3606 (2001). 

\bibitem{aya} B.Q., Wei, R. Vajtai, P.M. Ajayan, Appl. Phys. Lett. {\bf 79}, 1172 
(2001). 
\bibitem {md1}M. Di Ventra and S.T. Pantelides, 
Phys. Rev. B {\bf 61}, 16207 (2000); M. Di Ventra, S. T. Pantelides, and N. D. Lang, 
Phys. Rev. Lett. \textbf{88}, 046801 (2002).

\bibitem {avir}See, e.g., {\em Molecular Electronics II}, 
A. Aviram, M. Ratner, and V. Mujica eds., (NY Academy of Sciences, NY, 2002). 

\bibitem{miochem} M. Di Ventra, N.D. Lang, and S.T. Pantelides, 
Chemical Physics {\bf 281}, 189 (2002).

\bibitem {yan}A. I. Yanson, G. Rubio-Bollinger, H. E. van den Brom, N.
Agra\"{\i}t, and J. M. van Ruitenbeek, Nature, \textbf{395}, 783 (1998).

\bibitem {solid}R. S. Sorbello, \textit{Solid State Physics}, edited by H.
Ehrenreich and F. Spaepen (Academic Press, New York, 1998), Vol \textbf{51},
p. 159, and references therein; R. Landauer and J. W. F. Woo, Phys. Rev. B \textbf{10}, 1266 (1974); A. K. Das and R. Peierls, J. Phys. C \textbf{8}, 3348 (1975); 
L. J. Sham, Phys. Rev. B \textbf{12}, 3142 (1975).

\bibitem{prec} Note that this definition of direct force is not unique in literature (see 
e.g. Ref.~\onlinecite{solid}). 

\bibitem {langt}N. D. Lang, Phys. Rev. B \textbf{52}, 5335 (1995);
\textbf{49}, 2067 (1994); Z. Yang, A. Tackett, M. Di Ventra, Phys. Rev. B {\bf 66}, 041405 (2002). 

\bibitem {md2}M. Di Ventra and N. D. Lang, Phys. Rev. B \textbf{65}, 045402 (2002). 

\bibitem{prec1} Note that a misprint appears in the expression of the force in 
Refs.~\onlinecite{md1,md2,miochem}. A negative sign is missing in the definition of forces in these 
references. See M. Di Ventra, S.T. Pantelides, and N.D. Lang, Phys. Rev. Lett. {\bf 89}, 139902 (2002). 

\bibitem{prec3} Similar considerations explain the almost 
linear dependence of the force on the single-atom wire (see Fig.~\ref{fig1}(d)). In this 
case the extra charge on the atom is almost zero at any bias with consequent zero 
contribution from the discrete spectrum, and linear contribution from the continuum spectrum 
as a function of bias. 

\bibitem{prec2} It is interesting to note at this point that the sum of the continuum- and 
discrete-state contributions to the force is larger for the second atom from the right electrode at all biases. A similar result has been inferred by Brandbyge \textit{et al.} 
[M. Brandbyge, N. Kobayashi, and M. Tsukada, Phys. Rev. B \textbf{60}, 17064 (1999)] from 
the analysis of the voltage drop in gold wires, suggesting that this effect is material 
independent.

\bibitem{prec4} Using a tight-binding approach Todorov {\em at al.} have shown that in linear 
response the largest current-induced force is almost constant for wire lengths of 
three atoms or more.~\cite{tod1} We show here that this trend is 
also valid for the average force at any bias. 

\end{thebibliography}
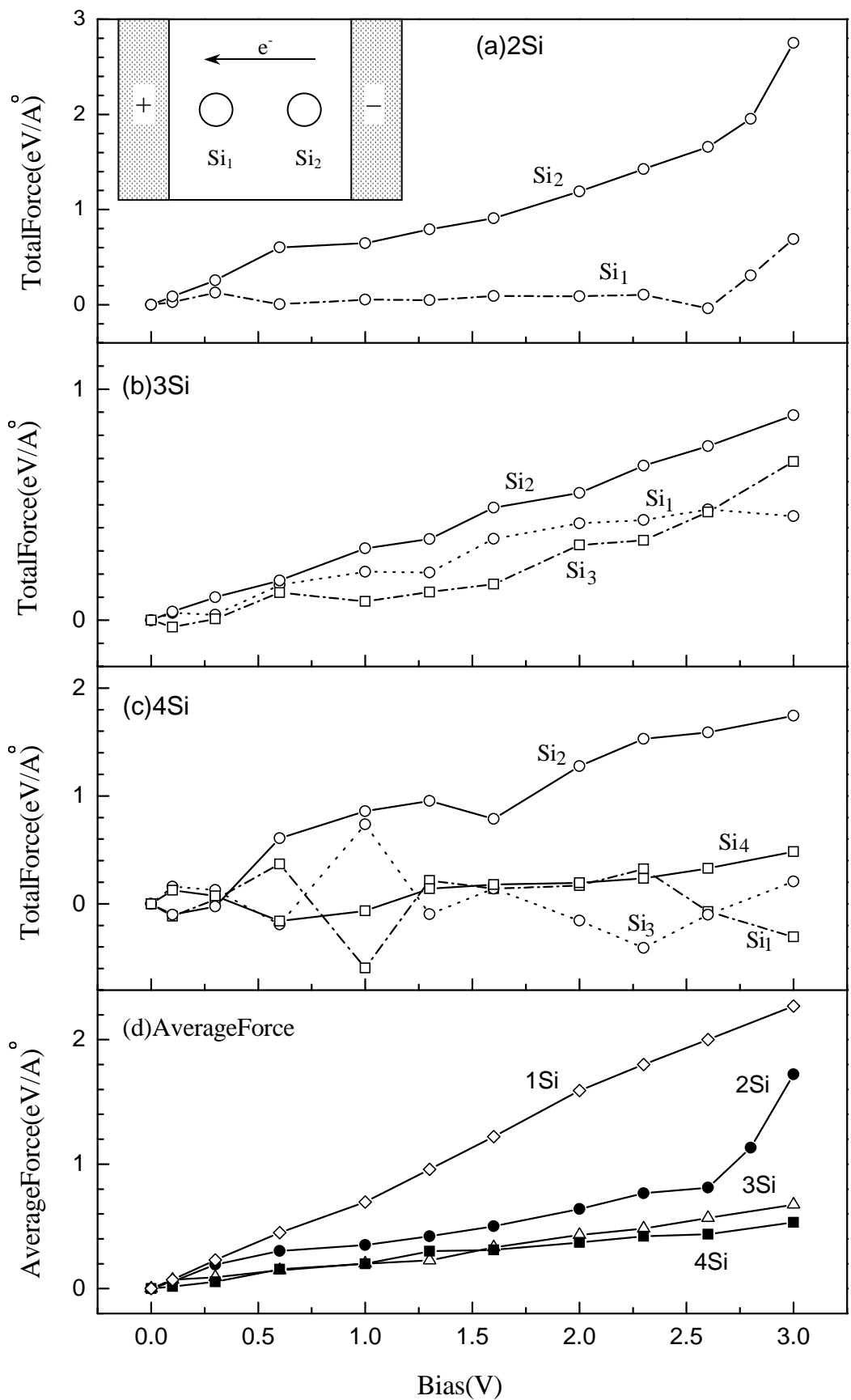
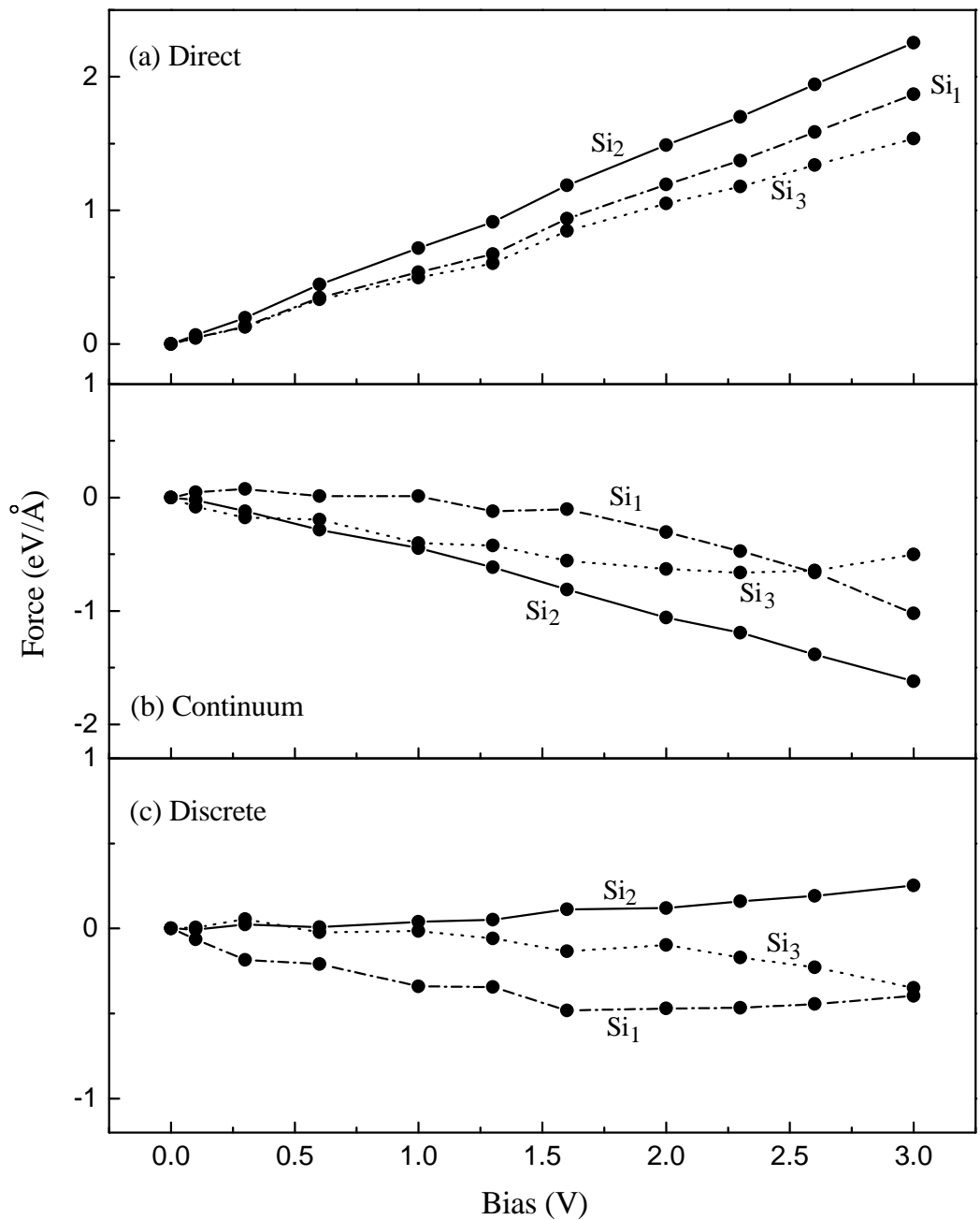
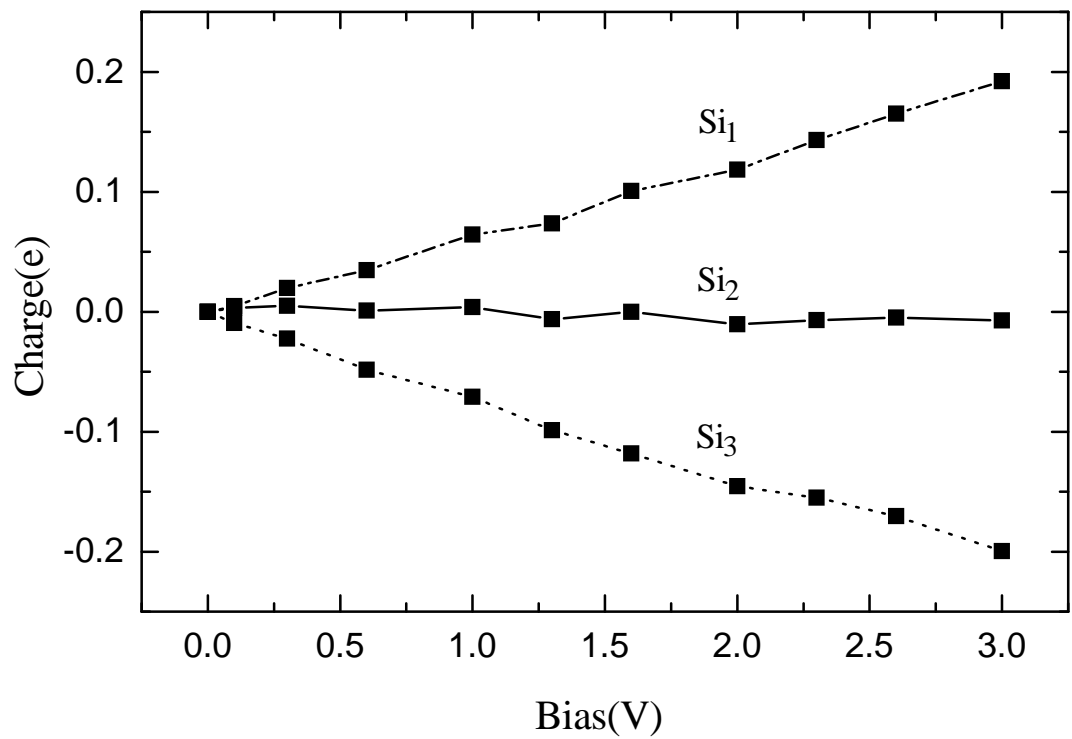
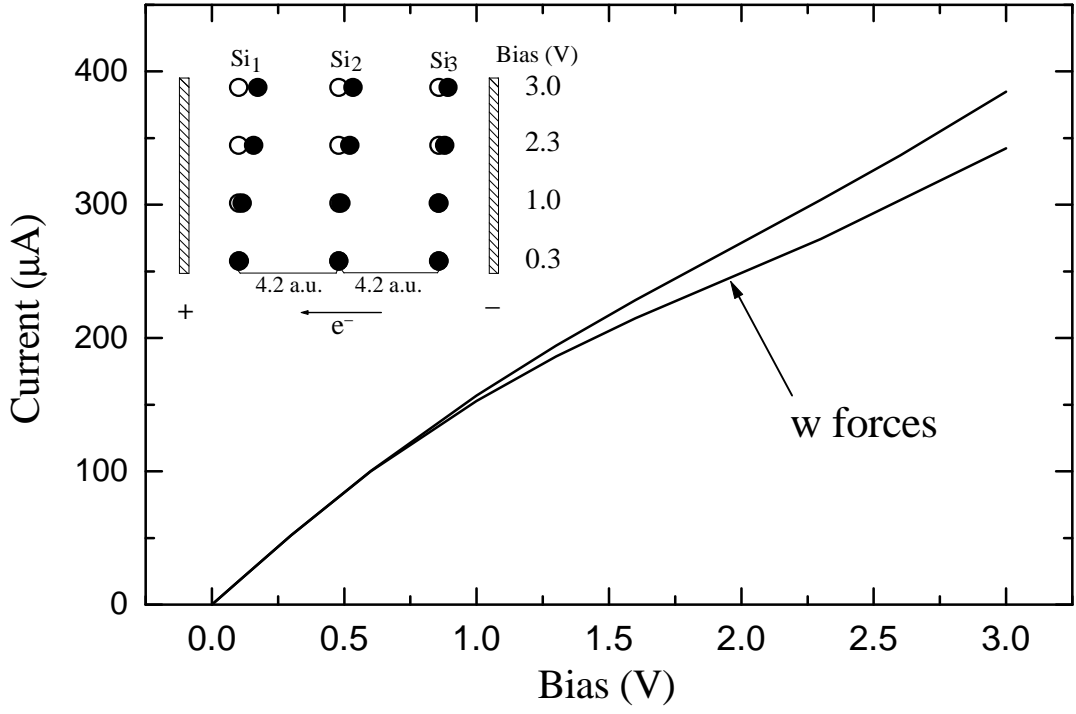
\end{document}